\documentclass[a4paper,12pt]{article}
\usepackage{graphicx}
\usepackage{times}
\usepackage{graphicx}
\usepackage{latexsym}
\usepackage{amsmath}
\usepackage{amssymb}
\usepackage{authblk}
\usepackage{setspace}
\begin{document}
\title{Development of Secular Instability in Different Disc Models of Black Hole Accretion}
\author[a]{Sankhasubhra Nag}
\author[b]{Deepika B Anand}
\author[c]{Ishita Maity}
\author[d]{Tapas K. Das}
\affil[a]{\small Department of Physics, Sarojini Naidu College for Women, Kolkata 700028}
\affil[b]{\small IISER, Pune, India}
\affil[c]{\small Louisiana State University, USA}
\affil[d]{\small HRI, Allahabad, India}
\date{}
\maketitle
\begin{abstract}
\noindent
Analytical treatment of black hole accretion generally presumes the stability of the stationary configuration. Various authors in the 
past several decades demonstrated the validity of such an assumption for inviscid hydrodynamic flow. 
Inviscid assumption is a reasonable approximation for low angular
momentum advection dominated flow in connection to certain
supermassive black holes at the
centres of the galaxies (including our own) fed from a number of stellar donors.
Introduction of a weak viscosity, however, may sometimes provide a
more detail understanding of the observed spectrum. Recently it has been demonstrated that introduction of small 
amount of viscosity in the form of quasi-viscous flow makes a stationary accretion disc -- where the geometric 
configuration of matter is described by axisymmetric flow in hydrostatic equilibrium -- unstable. 
We perform similar analysis for other disc models (for all three possible geometric configurations of 
matter) for quasi-viscous models under the post-Newtonian scheme. We introduced perturbations on the stationary 
flow solution particularly in standing wave form and studied their time evolution to observe whether 
they grow with time. Our analysis shows that same sort of secular instability exists in other disc models too. 
We further argued that with sufficiently low value of viscosity in the realistic astrophysical perspective,
the instability does not effectively jeopardize the stationary condition. 
\end{abstract}
\section{Introduction}
Astrophysical accretion is an important phenomena in connection to the
observational identifications of black hole candidates \cite{fkr02}. As the general MHD solution in this case is almost impossible to obtain analytically, people generally use hydrodynamical models to get an approximate analytical solution (see \cite{Chaudhury2006,swagata} and references therein). But in analytical analysis, generally the features of the stationary configuration are studied under different conditions such as equation of state, disc geometries, dynamical formalisms etc. But all those analysis implicitly presume the stability  of the stationary conditions. But recently some of the authors (e.g.~\cite{Bhattacharjee2007,Bhattacharjee2009}) demonstrated that with the introduction of a very small amount of viscosity may give rise to instabilities. But all the results available in the literature were worked out in a particular type of disc model, i.e. the model based on hydrostatic equilibrium along vertical equilibrium (henceforth to be referred as VE) and with adiabatic equation of state  only. We generalised these studies to all the disc models available in the literature (see \cite{Chakrabarti2001}).

\section{Basic Equations}
If some fluid is being accreted forming a disc like structure into some accretor, the rate of incoming mass into a thin ring shaped region of radius $r$ must be equal to the rate of increase of fluid mass in the region. Denoting the local surface mass density on the disc by $\Sigma$ and the inward radial velocity of the fluid by $v$, this consideration gives the continuity equation \cite{fkr02},
\begin{equation}
\dfrac{\partial}{\partial t}(\Sigma)+ \dfrac{1}{r}\dfrac{\partial}{\partial r}(\Sigma vr)=0
\label{conti}\end{equation}
where $ \Sigma = \rho H $, $\rho$ being the local density and $H$ being local disc half width.

The dependence of $H$ on radial distance $r$ varies in different disc models available \cite{Chakrabarti2001} in the literature. These are,\begin{subequations}\begin{align}
H &= H_0 &\textrm{for constant height disc (CH)},\\  H &= \Theta r,\;\;\;\; \Theta \textrm{ being a constant} &\textrm{for conical flow model (CM)},\\ H &=c_s(r)\sqrt{\frac{r}{\gamma\phi^{\prime}(r)}} &\textrm{for vertical equilibrium model (VE)},
\end{align}\end{subequations}
where, $c_s(r)=\sqrt{\partial P/\partial\rho}$ is the local acoustic speed with polytropic equation of state $P=K\rho^{\gamma}$ ($P$ being local pressure and $P=K\rho$ in isothermal cases).

Now for all the models $\rho H $ may be written in the form $ \rho^{1+\varepsilon} \bar{g}(r)$ for some constant $\epsilon$ and some function $\bar{g}(r)$, both of which vary from model to model.

For the disc geometry there must be azimuthal symmetry. But setting the Euler equation in lateral direction one gets the angular momentum Balance equation
\begin{equation}
\rho H \dfrac{\partial}{\partial t}(r^{2}\Omega)+ \rho v H \dfrac{\partial}{\partial r}(r^{2}\Omega) = \dfrac{1}{2 \pi r}\left( \dfrac{\partial G}{\partial r}\right)\label{ang}
\end{equation}
where $ G = 2 \pi \nu \Sigma r^{3} \dfrac{\partial \Omega}{\partial r} $ , $ \nu = \alpha c_{s} H $.

Euler equation in radial equation takes the form,
\begin{equation}
\dfrac{\partial v}{\partial t} + v \dfrac{\partial v}{\partial t} + \dfrac{1}{\rho} \dfrac{\partial P}{\partial r} + \Phi^{'}(r) - \dfrac{\lambda_{0}^{2}}{r^{3}}-\dfrac{2\alpha \lambda_{0}F}{r^{3}}=0\label{radial}
\end{equation} where,\begin{equation}
r^{2} \Omega = \lambda_{0} + \alpha \lambda_{0} F(c_{s},v,r)\label{scldang}
\end{equation}
Now, Eq.\eqref{conti} may be recast  as, 
\begin{equation}
\dfrac{\partial}{\partial t}(g_{1}(\rho))+ g_{2}(r) \dfrac{\partial}{\partial r}(f)=0\label{modcon}\end{equation} where
$
g_{1}(\rho)= \rho^{1+\varepsilon},
g_{2}(\rho)= \dfrac{1}{r \bar{g}(r)} \textrm{ and }
f( \rho ,v,r) = \rho vrH 
$
and Eq.\eqref{ang} may be written as (where $ f_{1} = \rho c_{s} H^{2} r^{3} $),
\begin{equation}
\dfrac{1}{v}\dfrac{\partial}{\partial t}(r^{2}\Omega)+\dfrac{\partial}{\partial r}(r^{2}\Omega) = \dfrac{\alpha}{f} \dfrac{\partial}{\partial r}(f_{1}\dfrac{\partial \Omega}{\partial r})\label{modang}
\end{equation}
Now putting  Eq.\eqref{scldang} in Eq.\eqref{modang} one gets,\begin{equation}
F(c_{s},v,r) = -2\left[\dfrac{f_{1}}{fr^{3}}+ \int \dfrac{f_{1}}{fr^{3}} \left(\dfrac{1}{f} \dfrac{\partial f}{\partial r} \right) dr \right].\label{Fsol}\end{equation}
\section{Stationary Solutions}
Stationary flow means for all the flow variables, $ \dfrac{\partial }{\partial t} \equiv 0 $ 
Hence Eq.'s \eqref{modcon} and \eqref{radial} take the form 
\begin{eqnarray}
\dfrac{\partial f}{\partial r} = 0 \label{stcon} \\
v \dfrac{\partial v}{\partial r} + \dfrac{1}{\rho} \dfrac{\partial P}{\partial r} + \Phi^{'}(r) - \dfrac{\lambda_{0}^{2}}{r^{3}}-\dfrac{2\alpha \lambda_{0}F}{r^{3}}=0\label{strad1}
\end{eqnarray}
\section{Introduction of Time Dependent Perturbation}
Now let us introduce a small time dependent perturbation on the stationary solution of the flow variables (denoted by suffix $0$) at each radial distance $r$, such that, 
\begin{subequations}
\begin{eqnarray}
v(r,t) = v_{0}(r)+ \tilde{v}(r,t), \\
\rho(r,t) = \rho_{0}(r)+ \tilde{\rho}(r,t), \\
f(r,t) = f_{0}(r)+ \tilde{f}(r,t), \\
g_{1}(r,t) = g_{10}(r)+ \tilde{g_{1}}(r,t), \\
g_{2}(r,t) = g_{20}(r)+ \tilde{g_{2}}(r,t), \\
F(r,t) = F_{0}(r)+ \tilde{F}(r,t).
\end{eqnarray}
To investigate the stability of the stationary solution, one has to find whether these perturbing terms grow with time or not. Substituting these expressions into the Eq.\eqref{modcon} and \eqref{radial} one gets an inhomogeneous wave equation,
\end{subequations} 
\begin{equation}
\dfrac{\partial } {\partial t} \left[ \dfrac{v_{0}}{f_{0}} \dfrac{ \partial \tilde{f}}{\partial t}\right] + \dfrac{\partial}{\partial t} \left[\dfrac{v_{0}^{2}}{f_{0}} \dfrac{\partial \tilde{f}}{\partial r} \right] + \dfrac{\partial}{\partial r}\left[\dfrac{v_{0}^{2}}{f_{0}}\dfrac{\partial \tilde{f}}{\partial t} \right] + \dfrac{\partial}{\partial r} \left[\dfrac{v_{0}}{f_{0}} \left( v_{0}^{2}-\dfrac{c_{s0}^{2}}{1+\varepsilon} \right)\dfrac{\partial \tilde{f}}{\partial r} \right]-\dfrac{2 \alpha \lambda_{0}^{2}}{r^{3}}\dfrac{\partial \tilde{F}}{\partial t}=0\label{wave}
\end{equation}
where,
\begin{equation}
\dfrac{\partial \tilde{F}}{\partial t} = \dfrac{c_{s0}}{\rho_{0} v_{0} r^{2}} \dfrac{(1+ \gamma + 4 \varepsilon)}{(1+ \varepsilon)} \dfrac{\partial \tilde{f}}{\partial r} + 2 \int \dfrac{\partial}{\partial r} \left( \dfrac{c_{s0}}{\rho_{0} v_{0}^{2}r^{2}}\right) \dfrac{\partial \tilde{f}}{\partial t} dr
\end{equation}

Now choosing a trial wave solution,
\begin{equation}
\tilde{f}(r,t) = g_{\omega}(r)e^{-i \omega t},
\end{equation} one gets may explore the behaviour of the spatial part.  Hence 
substituting the above in Eq.\eqref{wave}, the equation becomes,
\begin{equation}
\begin{split}
\omega^{2}v_{0}g_{\omega}^{2}+v_{0}^{2}i \omega \dfrac{d g_{\omega}}{dr} g_{\omega} + i\omega \dfrac{d}{dr}(v_{0}^{2}g_{\omega})g_{\omega} - g_{\omega} \dfrac{d}{dr}\left( v_{0} \left( v_{0}^{2} - \dfrac{c_{s0}^{2}}{1+ \varepsilon}\right) \dfrac{d g_{\omega}}{dr} \right) + &\\
 \dfrac{2 \alpha \lambda_{0}^{2} c_{s0}f_{0}g_{w}}{\rho_{0} v_{0} r^{5}}\left(\dfrac{1+ \gamma + 4 \varepsilon}{1+ \varepsilon} \right) \dfrac{dg_{\omega}}{dr} - \dfrac{4 \alpha \lambda_{0}^{2}i \omega f_{0}g_{\omega}}{r^{3}} \int \dfrac{d}{dr}\left( \dfrac{c_{s0}}{\rho_{0} v_{0}^{2}r^{2}}\right) g_{\omega} dr &= 0.\label{wv}
\end{split}
\end{equation}
The solution of this equation may be of two types. One is of standing wave type, where $\tilde{f}$ takes the form $A_{\omega}(r,t)\exp(-i\Omega t)$ where both $A_{\omega}(r,t)$ and $\Omega$ are real. The other is of travelling wave type where $\tilde{f}$ takes the form $A_{\omega}(r,t)\exp\left[i(s(r)-\omega t)\right]$, all of $A_{\omega}(r,t)$, $s(r)$ and $\omega$ being real. Here we will discuss the behaviour of the standing wave wave solutions and that of the travelling solutions will be reported elsewhere. 
\section{Results \& Comments}
Upon integrating the above equation so that the surface integral terms vanish we get finally something like
$
A \omega^{2}+B \omega+C = 0,
$
where
\begin{eqnarray*}
A = \int v_{0} g_{\omega}^{2} dr \;\;\;\;
B = - 4i \alpha \lambda_{0}^{2} \int \dfrac{g_{\omega} f_{0}}{r^{3}}\left[ \int g_{\omega} \dfrac{d}{dr} \left( \dfrac{c_{s0}}{\rho_{0} v_{0}^{2}r^{2}}\right) dr\right] dr \\
C = \int v_{0} \left(v_{0}^{2} - \dfrac{c_{s0}^{2}}{1+ \varepsilon} \right) \left(\dfrac{d g_{\omega}}{dr} \right)^{2} + 2 \alpha \lambda_{0}^{2} \int \dfrac{c_{s0}f_{0}}{\rho_{0} v_{0} r^{5}}\left( \dfrac{1+ \gamma + 4 \varepsilon}{1+ \varepsilon} \right)g_{\omega} \dfrac{d g_{\omega}}{dr} dr
\end{eqnarray*}

\begin{equation}
Re(-i \omega) = \alpha \left [ \int v_{0} g_{\omega}^{2} \xi (r) dr \right ] \left [ v_{0} g_{\omega}^{2} dr \right ]^{-1} \sim \alpha \xi (r)
\end{equation}
where 
\begin{equation}
\xi (r) = \dfrac{2 \lambda_{0}^{2}}{r^{3} g_{w}} \dfrac{g_{10}}{g_{20}} \int g_{w} \dfrac{d}{dr} \left( \dfrac{c_{s0} g_{10}^{\left( \dfrac{1+2 \varepsilon}{1+ \varepsilon} \right) }}{g_{20}^{2}f_{0}^{2}r^{2}} \right) dr
\end{equation}
Now after putting the specific  forms of $g_1$, $g_2$ and $\epsilon$, one gets for CH model $\xi (r) \sim r^{-2}$ while for CM $\xi (r) \sim r^{1}$ and for VE $\xi (r) \sim r^{5/2}$; which readily implies  that at large radial distance for CM \& VE the perturbation will rapidly grow with time making the disc unstable but for CH it will diminish with radial distance making the disc practically stable for large amount of time. Moreover in all  cases with sufficiently smaller value of $\alpha$, the instability may grow too slowly to sustain the disc for large enough time in comparison with the  lifespan of the disc otherwise realizable.

\end{document}